\documentstyle{article}
\begin{document}
\baselineskip12pt

\begin{center}

{\Large\bf Spacetime Measurements in Kaluza-Klein Gravity}\\

\vspace{.25in} {\footnotesize Hongya Liu} \\ {\footnotesize \it Department of Physics,
Dalian University of Technology, Dalian, 116024, P.R. China}\\
\vspace{.10in} {\footnotesize Bahram Mashhoon} \\ {\footnotesize \it Department of
Physics and Astronomy, University of Missouri-Columbia}\\ {\footnotesize \it Columbia,
Missouri 65211, USA}\\

\end{center}

\begin{abstract}  We extend the classical general relativistic theory of measurement
to include the possibility of existence of higher dimensions.  The intrusion of these
dimensions in the spacetime interval implies that the inertial mass of a particle in
general varies along its worldline if the observations are analyzed assuming the existence of only the four spacetime
dimensions.  The variations of mass and spin are explored in a simple 5D Kaluza-Klein model.

\begin{itemize}
\item  PACS numbers: 04.20.Cv, 04.50.+h 
\end{itemize}
\end{abstract}

\baselineskip17pt
\vspace{.10in}
\noindent{\Large \bf 1 Introduction}

\vspace{.10in}

The most basic measurements of a physical observer are those of time and space.  In
principle, such observations may involve an atomic clock for local temporal
measurements and three independent spatial axes for the characterization of space in
the neighborhood of the observer.  In standard general relativity, the observer
carries an orthonormal tetrad frame along its worldline and physical observables are
scalars that are obtained as the projections of tensors that correspond to various
physical variables upon the local tetrad frame of the observer.  In general relativity,
just as in Newtonian physics, the observer can determine, via local measurements, the
acceleration of its local frame.  This acceleration could be in the form of the
translational acceleration of the observer as well as the rotation of its local
spatial frame.  Theoretically, a set of three ideal orthogonal torque-free gyroscopes
can provide a nonrotating (i.e. Fermi-Walker transported) spatial frame along the
path of the observer.  Thus in 4D spacetime a free test observer can carry a
nonrotating orthonormal frame that is parallel transported along its geodesic worldline.  In
view of the possibility of existence of higher dimensions, it is worthwhile to
examine how the theory of measurement in general relativity would have to be extended
if higher dimensions intrude into the spacetime arena.  This intrusion is expected in any
realistic higher-dimensional physical theory [1, 2]; nevertheless, the interpretation
of observational data currently involves only the standard four spacetime dimensions. 

In this paper, we explore the extension of the classical general relativistic theory
of measurement to the Kaluza-Klein theory by studying some of the main observational
consequences of the dependence of the spacetime metric upon extra dimensions. 
For instance, we show in section 2 that the mass of a test particle in general varies
along its worldline if the motion of matter is not wholly confined to the four
spacetime dimensions.  This circumstance is expected in realistic higher-dimensional
theories [1, 2].  Our physical considerations are motivated by the fact that
experimental data are routinely analyzed assuming the existence of only the four
spacetime dimensions.  In section 3, we show that an initially orthonormal frame does
not in general remain orthonormal along the worldline of a test observer.  To render
these results explicit, we consider a concrete 5D model in section 4 and explore its
physical consequences.  Section 5 contains a brief discussion of our results. \\
\vspace{.10in}

\noindent{\Large\bf 2 Variation of the Inertial Mass}

\vspace{.10in}

Imagine that the universe can be described in terms of $4+N$ dimensions with 
$N\geq 1$ and the 4D spacetime part, which is embedded in the $4+N$ manifold, has a
metric of the form 
\begin{equation} ds^2=g_{\mu \nu }(x,y)dx^\mu dx^\nu .  \label{ds}
\end{equation} Here $x$ stands for the spacetime coordinates and $y$ stands for the
extra dimensions $\left( y^1,...,y^N\right) $ that are in general reflected in the
spacetime metric. Greek indices run from 0 to 3.  The path of a test particle in the $4+N$ dimensional manifold
involves the 4D velocity $u^\alpha =dx^\alpha /ds$, where $s$ is the proper time
along the path such that 
\begin{equation} g_{\mu \nu }(x,y)u^\mu u^\nu =1.  \label{ds/ds}
\end{equation} Differentiating (2) with respect to $s$, we find that 
\begin{equation} g_{\mu \nu ,\alpha }u^\alpha u^\mu u^\nu +\sum_{i=1}^N\frac{\partial
g_{\mu
\nu }}{\partial y^i}\frac{dy^i}{ds}u^\mu u^\nu +2g_{\mu \nu }u^\mu \frac{du^\nu }{ds}=0.
\end{equation} This relation may be written in the form 
\begin{equation}
\left( g_{\mu \nu ,\alpha }+g_{\mu \alpha ,\nu }-g_{\nu \alpha ,\mu }\right) u^\alpha
u^\mu u^\nu +2g_{\mu \nu }u^\mu \frac{du^\nu }{ds}=-\sum_{i=1}^N%
\frac{\partial g_{\mu \nu }}{\partial y^i}\frac{dy^i}{ds}u^\mu u^\nu .
\label{ds/ds-1}
\end{equation} Using the fact that the 4D connection is given by 
\begin{equation}
\Gamma _{\alpha \beta }^\nu \left( x,y\right) =\frac 12g^{\nu \eta }\left( g_{\eta
\alpha ,\beta }+g_{\eta \beta ,\alpha }-g_{\alpha \beta ,\eta }\right) ,
\end{equation} we can write equation (4) as 
\begin{equation} 2g_{\mu \nu }u^\mu \left( \frac{du^\nu }{ds}+\Gamma _{\alpha \beta
}^\nu u^\alpha u^\beta \right) =-\sum_{i=1}^N\frac{\partial g_{\mu \nu }}{\partial
y^i}\frac{dy^i}{ds}u^\mu u^\nu .  \label{ds/ds-2}
\end{equation} The 4D acceleration of the particle is defined by 
\begin{equation} A^\mu =\frac{du^\mu }{ds}+\Gamma _{\alpha \beta }^\mu u^\alpha
u^\beta ,
\label{A}
\end{equation} so that equation (6) implies 
\begin{equation} u_\mu A^\mu =-\frac 12\sum_{i=1}^N\frac{\partial g_{\mu \nu
}}{\partial y^i}%
\frac{dy^i}{ds}u^\mu u^\nu .  \label{uA}
\end{equation}

If $g_{\mu \nu }$ does not depend on the extra dimensions, then equation (8) becomes
$u_\mu A^\mu =0$, as expected. However, in the higher-dimensional theories
$g_{\mu \nu }$ may depend on $y$ and $y$ may vary with respect to $s$. So the
right-hand side of equation (8) may not vanish. Let us note that $u^{\mu}$ is a
timelike vector, so $u_{\mu} A^{\mu} \neq 0$ indicates that there may be a {\it
timelike} component of acceleration in $A^\mu $ in higher-dimensional theories. This
is an extraordinary result, since all known basic 4D forces are spacelike and lead to
accelerations that are orthogonal to the 4D velocity of the particle [3].  It turns
out that the most natural way to incorporate a {\it timelike} acceleration into 4D
physics is to assume that the ``invariant'' inertial mass of the test particle varies
along its worldline.

Experimental data are reduced and interpreted at present assuming that the most
general force law for the motion of a test particle may be written classically as 
\begin{equation}
\frac{Dp^\mu }{ds}\equiv \frac{dp^\mu }{ds}+\Gamma _{\alpha \beta }^\mu u^\alpha
p^\beta =F^\mu ,  \label{dp/ds}
\end{equation} where $p^\mu \equiv mu^\mu $ is the momentum of the particle and $F^\mu $ consists of all forces
acting on the particle arising from the known fundamental interactions. In the rest
frame of the particle, we expect all forces acting on the particle to be 3D vectors;
therefore, $u_\mu F^\mu =0$. This relation together with the force law (9),
$Dp^\mu /ds=\left( dm/ds\right) u^\mu +mA^\mu=F^\mu$, implies that 
\begin{equation}
\frac 1m\frac{dm}{ds}=\frac 12\sum_{i=1}^N\frac{\partial g_{\mu \nu }}{%
\partial y^i}\frac{dy^i}{ds}u^\mu u^\nu ,  \label{dm/ds}
\end{equation} where we have used equation (8). That is, the simplest interpretation
of equation (8) in terms of 4D physics is to assume that the invariant ``rest'' mass
of the particle may vary with respect to its proper time $s$ due to the existence of
higher dimensions. Conversely, the observation of such a basic variation would
indicate the presence of an extra {\it timelike} acceleration and this could come
about precisely because of the intrusion of the extra dimensions into the 4D physics
as indicated by equations (1) and (8).

Let us note that equation (10) may be expressed as 
\begin{equation}
\delta \left( m^2\right) =\left[ \sum_{i=1}^N\frac{\partial g_{\mu \nu }(x,y)%
}{\partial y^i}\delta y^i\right] p^\mu p^\nu ,  \label{delta-m}
\end{equation}
where 
\begin{equation} m^2=g_{\mu \nu }(x,y)p^\mu p^\nu .  \label{m^2}
\end{equation} Equations (11) and (12) can be used to determine the variable inertial
mass in higher-dimensional theories.

The acceleration $A^\mu$ and the variation of the extra coordinates of the particle
along its worldline are determined by the equation of motion of the theory. In
principle, it is possible that $y$ can vary in just such a way as to render $dm/ds=0$
in equation (10). This would, of course, require rather special circumstances;
therefore, we pursue in this paper the general situation in which $dm/ds\neq 0$.\\

\vspace{.10in}
\noindent{\Large\bf 3 Variation of the Spin}
\vspace{.10in}

Let us first consider an ideal gyroscope represented by the spin vector
$\sigma^{\mu}$ within the context of classical general relativity.  We may imagine in
our classical model that we are dealing here with the limiting case of an ideal
gyroscope with its magnitude of spin given by $\sigma = I\omega$ in a certain
``rest'' frame of the system.  Here $I$ is the proper moment of inertia of the
gyroscope and $\omega$ is its angular speed of rotation with respect to the {\it
proper time} of the ``pointlike'' spinning particle.  The general relativistic theory
is based on the Mathisson-Papapetrou equations for a ``pole-dipole'' test particle. 
It follows from detailed considerations [4, 5] of the classical theory of such ideal
spinning ``point'' particles that $\sigma _\mu u^\mu =0$.  Moreover, $\sigma^{\mu}$ is nonrotating, i.e. 

\begin{equation}
\frac{D\sigma^{\mu}}{ds} = -(\sigma_{\alpha}A^{\alpha})u^{\mu}\;\;\; ,
\end{equation}

\noindent so that $\sigma_{\alpha}u^{\alpha}$ and $\sigma ^2=-g_{\mu \nu }\sigma ^\mu \sigma ^\nu$ are constants
along the
worldline of the particle.

Consider next the possibility that the spacetime metric could depend upon higher
dimensions.  In this case, one can extend the treatment of section 2 to demonstrate
that 

\begin{equation} Dg_{\mu\nu} = \sum^{N}_{i=1}\; \frac{\partial g_{\mu\nu}}{\partial
y^i}\; dy^i\;\;\; .
\end{equation}

\noindent Using this result, it is straightforward to show that 

\begin{equation}
\frac{d}{ds}\;(\sigma_{\mu}u^{\mu}) = T_{\mu} u^{\mu} + A_{\mu}\sigma^{\mu} +
\sum^{N}_{i=1}\;
\frac{\partial g_{\mu\nu}}{\partial y^i}\;\;\frac{dyi}{ds}\;\sigma^{\mu}u^{\nu}
\end{equation}

\noindent and 

\begin{equation}
\frac{d}{ds}\;(\sigma^2) = -2 T_{\mu}\sigma^{\mu} - \sum^{N}_{i=1}\;\frac{\partial
g_{\mu\nu}}{\partial y^i}\;\;\frac{dyi}{ds}\;\sigma^{\mu}\sigma^{\nu}\;\;\; ,
\end{equation}

\noindent where $T^{\mu} = D\sigma^{\mu}/ds$ is the torque and $A^{\mu}$ is the
translational acceleration (7) of the particle.  The vectors $A^{\mu}$ and $T^{\mu}$
must be determined from the higher-dimensional theory under consideration.  It
follows from the extension of these results to the axes of a spatial frame carried by
an observer that an initially orthonormal frame will not in general remain
orthonormal in the course of time.  In particular, the magnitude of spin can change
with time.  To illustrate how this could come about, we now turn to a simple
Kaluza-Klein gravitational model.\\

\vspace{.10in}
\noindent{\Large\bf 4 A 5D Model}
\vspace{.10in}

We consider a Kaluza-Klein model with one extra noncompactified spacelike dimension $y$ such that the
5D metric is given by

\begin{equation} dS^2 = \hat{g}_{AB}dx^{A}dx^{B} = ds^{2} - dy^{2}\;\;\; ,
\end{equation}

\noindent where $x^{A}=(x^{\mu}, y)$ and $\hat{g}_{AB}$ satisfies the 5D vacuum field
equations
$\hat{R}_{AB}=0$.  It is natural to assume in this theory that a test particle
follows the 5D geodesic equation

\begin{equation}
\frac{d^2x^A}{dS^2}+\widehat{\Gamma }_{BC}^A\;\frac{dx^B}{dS}\;\frac{dx^C}{dS}=0\;\; ,
\end{equation}

\noindent so that the 5D velocity vector $U^{A} = dx^{A}/dS$ is parallel transported
along the path.  Moreover, we expect that the 5D spin vector $\Sigma^{A}$ of an ideal
test gyroscope would also be parallel transported along its path

\begin{equation}
\frac{d\Sigma^A}{dS} + \hat{\Gamma}^{A}_{BC}U^{B}\Sigma^{C} = 0\;\;\; .
\end{equation}

\noindent To connect equations (18)-(19) with our spacetime variables in previous
sections, we note that 

\begin{equation} u^{\mu} = \left(\frac{dS}{ds}\right)U^{\mu}\;\;\; , \;\;\;
\sigma^{\mu} =
\left(\frac{dS}{ds}\right)\Sigma^{\mu}\;\;\; .
\end{equation}

\noindent The relationship between 4D and 5D velocities follows simply from the
definitions, while the corresponding relation for spin is expected by analogy.  

The Kaluza-Klein field equations $\hat{R}_{AB}=0$ can be reduced to certain constraint equations together with the
4D gravitational field equations of standard general relativity with an effective energy-momentum tensor as the
source of the gravitational field.  This is consistent with recent investigations [6] based on Campbell's theorem
that an $n$-dimensional Riemannian space can be locally embedded in a Ricci-flat $(n+1)$-dimensional Riemannian space
[7].

To proceed further, we need an explicit solution of the field equations.  It is
possible to show that with

\begin{equation} g_{\mu\nu}(x, y) = \frac{y^2}{L^2}\; \tilde{g}_{\mu\nu}\;(x)
\end{equation}

\noindent in the spacetime interval (1), $\hat{g}_{AB}$ in equation (17) is a
solution of
$\hat{R}_{AB} = 0$, provided that $\tilde{g}_{\mu\nu}(x)$ is {\it any} source-free
solution of general relativity with a cosmological constant $\tilde{\Lambda} =
3/L^2$.  Here $L$ is a constant length.  If
$\tilde{g}_{\mu\nu}(x)$ is the de Sitter solution, then $\hat{R}_{ABCD} = 0$ and the
5D metric is flat.  However, for more complicated Einstein spaces such as the Kerr-de
Sitter solution the 5D manifold is curved.  Let us note that for $g_{\mu\nu}(x, y)$,
the 4D Ricci tensor is then given by $R_{\alpha\beta} = -3y^{-2} g_{\alpha\beta}$. 
Thus the spacetime metric can be interpreted as a source-free solution of general
relativity with a cosmological ``constant'' $\Lambda = 3/y^2$.  Various aspects of
the gravitational model under consideration here have been explored in a number of
publications [8].  A recent discussion of variable cosmological ``constant'' is
contained in [9].

With a spacetime metric of the form (21), equation (18) reduces to $Du^{\alpha}/ds =
A^{\alpha}$ with 

\begin{equation} A^{\alpha} = -\frac{1}{y}\;\frac{dy}{ds}\;\;u^{\alpha}
\end{equation}

\noindent and

\begin{equation}  y\frac{d^2y}{ds^2}-\left( \frac{dy}{ds}\right) ^2+1=0\;\; . 
\label{Eq-y}
\end{equation}

\noindent Equation (10) --- or, equivalently, equation (22) --- implies that $m=\lambda
y$, where
$\lambda$ is a constant.  Moreover, equation (23) can be solved by studying the variation of $dy/ds$ with
respect to $y$.  The result is 

\begin{equation}
\left( \frac{dy}{ds}\right) ^2+\frac K{L_0^2}\,y^2=1\;\;,
\end{equation}
 
\noindent where $K=0,\pm 1$ and $L_0$ is a constant length. It follows that 

\begin{equation}
\pm y=\left[ 
\begin{array}{ll} L_0\sin \left( \frac{s-s_0}{L_0}\right)  & \qquad {\rm for}\:K=+1,
\\  s-s_0 & \qquad {\rm for}\:K=0, \\  L_0\sinh \left( \frac{s-s_0}{L_0}\right)  &
\qquad {\rm for}\:K=-1.
\end{array}
\right. 
\end{equation} 

If we now substitute for $y$ from equation (25) in $dS^2 = ds^2-dy^2$, we find that
$dS^2$ is greater than, equal to, or less than zero for $K = +1, 0, -1$,
respectively.  In general relativity, we know that the worldline of a massive
(massless) particle should have $ds^2 > 0$ ($ds^2 =0 $).  Extending this requirement
from 4D to 5D, we choose $dS^2 > 0$ ($dS^2 = 0$) for the motion of a massive
(massless) particle.  This implies that $K = +1$ and $K = 0$ in equation (25) hold
for massive and massless particles, respectively.  It follows from this
interpretation that for $K = 0$ we have $ds = 0$, so that
$s = s_0$ and hence $y = 0$.  Let us note that this interpretation is consistent with
equation (25), since for $s - s_0 \rightarrow 0$ the timelike and spacelike cases $K
= \pm 1$ reduce to the null case $K = 0$ regardless of the value of $L_0$.  Moreover,
for $s - s_0 \rightarrow 0, y \rightarrow 0$, and hence
$m = \lambda y \rightarrow 0$, so that lightlike propagation occurs only for a
massless particle.  Thus null rays propagate only in the 4D spacetime part of the 5D
manifold.

The case of a massless particle is important for the treatment of light propagation;
therefore, it is necessary to show in detail how the null geodesic case comes about
as a limiting case of the motion of a massive particle for $m \rightarrow 0$.  This
involves a standard limiting procedure in general relativity that will be adapted
here to the situation at hand.  Let us note that for a massive particle the momentum
$p^\alpha = mu^\alpha$ is covariantly constant along its 4D worldline.  Here $m =
\lambda y$ and $y$ is given by equation (25).  Thus we choose a new variable $z$
defined by $s - s_0 = \varepsilon z$, where $\varepsilon, 0 < \varepsilon \ll 1$, is
a constant such that $\varepsilon\rightarrow 0$ in the massless limit.  If we now let
$q = \pm \lambda^{-1}$ ${\rm ln}\; z$ be an affine parameter along the worldline, then it is
simple to show that for $\varepsilon\rightarrow 0, p^\alpha \rightarrow k^\alpha =
dx^\alpha/dq$, which is tangent to a null ray, and that the equation of motion
reduces in this limit to the null geodesic equation $Dk^\alpha/dq = 0$.

For a massive particle we have from the first equation in (25)

\begin{equation} 
m = m_{\rm{max}}\sin
\frac{s-s_0}{L_0}\;\;\;\;,\;\;\;\;\frac{1}{m}\;\frac{dm}{ds}\;=\;\frac{1}{L_0}\;{\rm
cot}\;\frac{s-s_0}{L_0}\;\; ,
\end{equation}

\noindent where $m_{\rm{max}} = \pm\lambda L_0$.  The present upper limit on
$|m^{-1}(dm/ds)|$ is of order
$10^{-12}$/yr, so that $L_0$ must be a sufficiently large cosmic length to render
equation (26) compatible with observation [10].

Let us now explore the implications of equation (19) for the motion of the gyro
axis.  We find that
$D\sigma^{\mu}/ds = T^{\mu}$, where the torque is given by 

\begin{equation} T^{\mu} =
-\;\frac{1}{y}\left(\frac{dS}{ds}\right)\;\Sigma^{4}\;u^{\mu}
\end{equation}

\noindent and the variation of $\Sigma^4$ is governed by

\begin{equation}
\frac{d\Sigma^4}{ds} +
\frac{1}{y}\;\left(\frac{ds}{dS}\right)\;\sigma_{\alpha}u^{\alpha} = 0\;\; .
\end{equation}

\noindent These equations are consistent with the fact that $\Sigma_A U^A$ and
$\Sigma_A\Sigma^A$ are constants along the path.  For instance,

\begin{equation}
\left(\frac{ds}{dS}\right)^{2}\;\sigma^{2} + \left(\Sigma^{4}\right)^{2} =
\Sigma^2_0\;\;\; ,
\end{equation}

\noindent where $\Sigma_{0} = (-\Sigma_{A}\Sigma^{A})^{1/2}$ is the constant
magnitude of the 5D spin vector.  One can now compute the variation of
$\sigma_{\alpha}u^{\alpha}$ and $\sigma^2$ along the path using equations (15) and
(16).  In particular, we find that $\sigma_{\alpha}u^{\alpha} = C_0 +
C_1{\rm cos}\; \theta$ and

\begin{equation}
\Sigma^4 = \xi (C_0\;{\rm cot}\; \theta + C_1\;{\rm csc}\;\theta)\;\;\; ,
\end{equation}

\noindent where $\theta = (s - s_0)/L_0$ and $\xi^{2}=1$ (i.e. $\xi$ is either $+1$ or $-1$).  Here $C_0$ and $C_1$
are constants of integration.  To simplify matters, let us assume that $C_0 = \Sigma_A U^A$
vanishes along the path, generalizing the standard 4D constraint for a ``pointlike'' gyroscope.  Moreover, we set
$C_1 = 0$; then,
$T^{\mu} = 0$ and $\sigma_{\alpha}u^{\alpha} = 0$.  It follows that one can choose
initial conditions such that $\sigma^{\mu}$ is nonrotating along the path, since
$\sigma_{\mu}A^{\mu} = 0$ follows from our assumptions.  However, the magnitude of
spin would still vary in accordance with equation (29), i.e. $\sigma =
\lambda^{\prime}y$, where
$\lambda^{\prime}\neq 0$ is a constant.  Hence the ``pole-dipole'' particle's mass
and spin both vary along its trajectory in such a way that
$\sigma/m$ remains constant.\\

\vspace{.10in}
\noindent{\Large\bf 5 Discussion}
\vspace{.10in}

A preliminary analysis of the basic spacetime measurements in higher-dimen\-sional
gravity theory reveals that the inertial mass and spin of an ideal classical
``pointlike'' gyroscope may vary along its worldline.  These results could be
significant in Kaluza-Klein cosmology as well as the search for extra dimensions.

Our discussion has been primarily concerned with classical general relativity;
however, the results are expected to be of more general validity.  In fact, the
intrusion of the extra dimensions in the spacetime domain implies that it is not
possible in general to reduce the spacetime metric to the Minkowski form at an
arbitrary event in spacetime.  This would indicate a breakdown of some of the basic
concepts of standard relativistic physics; for instance, the inertial mass and the
magnitude of spin are invariant constants that characterize the irreducible unitary
representations of the Poincar\'{e} group, but could now become variables.  We have
explored this circumstance in this paper within the eikonal approximation of the
classical theory of gravitation; nevertheless, our treatment could be extended to other physical quantities such as
the variation of the phase of a wave.

\vspace{.25in}

H.~L.~acknowledges the support of the National Natural Science Foundation of China
(grant no. 19975007).

\newpage

\noindent{\Large\bf References}

\begin{description}

\item{[1]}  T. Kaluza, Sitz. Preuss. Akad. Wiss. {\bf 33}, 966 (1921); O. Klein, Z.
Phys. {\bf 37}, 895 (1926); T. Appelquist, A. Chodos and P. G. O. Freund, {\it Modern
Kaluza-Klein Theories }(Addison-Wesley, Menlo Park, CA, 1987); J. M. Overduin and P. S.
Wesson, Phys. Rep. {\bf 283}, 303 (1997); P. S. Wesson, {\it Space, Time, Matter: Modern
Kaluza-Klein Theory }(World Scientific, Singapore, 1999).

\item{[2]} C. Cs\'{a}ki, M. Graesser, C. Kolda and J. Terning,
Phys. Lett. B \textbf{462}, 34 (1999); J. M. Cline, C. Grojean and
G. Servant, Phys. Rev. Lett. {\bf 83}, 4245 
(1999); L. Randall and R. Sundrum, Phys. Rev. Lett. {\bf 83}, 4690
(1999); R. Maartens, \emph{Cosmological Dynamics on the Brane},
hep-th/0004166. 
 
\item{[3]} B. Mashhoon, P. S. Wesson and H. Liu, Gen. Rel. Grav. {\bf 30}, 555
(1998); P. S. Wesson, B. Mashhoon, H. Liu and W. N. Sajko, Phys. Lett. B {\bf 456},
34 (1999).

\item{[4]} F. A. E. Pirani, Acta Phys. Polon. {\bf 15}, 389 (1956).

\item{[5]} B. Mashhoon, J. Math. Phys. {\bf 12}, 1075 (1971); Ann. Phys. (NY) {\bf
89}, 254 (1975).

\item{[6]} C. Romero, R. Tavakol and R. Zalatdinov, Gen. Rel. Grav. {\bf 28}, 365 (1996); J. E. Lidsey, C. Romero,
R. Tavakol and S. Rippl, {\it Class. Quantum Grav.} {\bf 14}, 865 (1997).

\item{[7]} J.E. Campbell, {\it A Course of Differential Geometry} (Clarendon Press, Oxford, 1926).

\item{[8]} B. Mashhoon, H. Liu and P. S. Wesson, Phys. Lett. B {\bf 331}, 305 (1994);
H. Liu and B. Mashhoon, Ann. Physik {\bf 4}, 565 (1995); B. Mashhoon, H. Liu and P.S. Wesson, in {\it Proc. Seventh
Marcel Grossman Meeting on General Relativity} (Stanford, 1994), edited by R. T. Jantzen and G. Mac Keiser (World
Scientific, Singapore, 1996), p. 333; P. S. Wesson, B. Mashhoon and Hongya Liu, Mod. Phys. Lett. A {\bf 12}, 2309
(1997).

\item{[9]} J. M. Overduin and F. I. Cooperstock, Phys. Rev. D {\bf 58}, 043506 (1998).

\item{[10]} I. I. Shapiro, in {\it Quantum Gravity and Beyond --- Essays in Honor of
Louis Witten,} edited by F. Mansouri and J. J. Scanio (World Scientific, Singapore,
1993), p. 180; J. D. Bekenstein, Phys. Rev. D {\bf 15}, 1458 (1977).

\end{description}
\end{document}